\documentclass[
preprint,longbibliography,
superscriptaddress,
amsmath,amssymb,
aps,
pra,
]{revtex4-2}

\usepackage{graphicx}
\usepackage{amsmath}
\usepackage{color}
\usepackage[normalem]{ulem}
\usepackage[colorlinks, linkcolor= blue, citecolor = blue, urlcolor=blue]{hyperref}
\newcommand\beq{\begin{equation}}
	\newcommand\eeq{\end{equation}}

\begin{document}

\title{Spin-controlled photonics via temporal anisotropy}
%\subtitle{Insert subtitle if needed}

\author{Carlo Rizza} 
\thanks{These two authors contributed equally}
\affiliation{Department of Physical and Chemical Sciences, University of L'Aquila, Via Vetoio 1, I-67100 L'Aquila, Italy}
\email{carlo.rizza@univaq.it}

\author{Giuseppe Castaldi}
\thanks{These two authors contributed equally}
\affiliation{University of Sannio, Department of Engineering, Fields \& Waves Lab, Benevento, I-82100, Italy} 

\author{Vincenzo Galdi} \email{vgaldi@unisannio.it}
\affiliation{University of Sannio, Department of Engineering, Fields \& Waves Lab, Benevento, I-82100, Italy}

\begin{abstract}	
Temporal metamaterials, based on time-varying constitutive properties, offer new exciting possibilities for advanced field manipulations. In this study, we explore the capabilities of anisotropic temporal slabs, which rely on abrupt changes in time from isotropic to anisotropic response (and vice versa). Our findings show that these platforms can effectively manipulate the wave-spin dimension, allowing for a range of intriguing spin-controlled photonic operations. We demonstrate these capabilities through examples of spin-dependent analog computing and spin-orbit interaction effects for vortex generation. These results provide new insights into the field of temporal metamaterials, and suggest potential applications in communications, optical processing and quantum technologies.
\end{abstract}


\maketitle

%%%%%%%%%%%%%%%%%%%%%%%%%%%%%%%%%%%%%%
\section{Introduction}
%%%%%%%%%%%%%%%%%%%%%%%%%%%%%%%%%%%%%%

In recent years, there has been mounting interest in ``temporal'' and ``space-time'' metamaterials \cite{Engheta:2021mw,Caloz:2020sm1,Caloz:2020sm2}. These are artificial materials where the  spatial modulation of the constitutive parameters is replaced by (or combined with) {\em time-varying} properties. This field of research, which has roots in longstanding theoretical foundations \cite{Morgenthaler:1958vm,Oliner:1961wp,Felsen:1970wp,Fante:1971to}, has been spurred by the emergence of new physical concepts such as ``time crystals'' \cite{Wilczek:2012qt,Sacha:2018tc}  and major technological advances in rapidly reconfigurable material constituents across the electromagnetic (EM) spectrum \cite{Kamaraju:2014sc,Alam:2016lo,Kord:2019ci,Khurgin:2020af,Zhou:2020bf}. 

Accessing the temporal dimension offers the potential for advanced spatial-spectral field manipulations and for surpassing fundamental limitations of linear, time-invariant systems \cite{Hayran:2022cf}. Accordingly, a broad variety of concepts relying on space-time analogies have been put forward, ranging from relatively simple ideas such as temporal boundaries \cite{Xiao:2014ra}, interfaces \cite{Stefanini:2022ti}  and slabs \cite{Ramaccia:2020lp}, to 
more sophisticated ones including gratings \cite{Galiffi:2020wa,Taravati:2019gs}, filters \cite{Ramaccia:2021tm,Castaldi:2022he}, photonic time crystals \cite{Romero:2016tp,Lustig:2018ta,Lyubarov:2022ae}, antireflection coatings \cite{Shlivinski:2018bt,Pacheco:2020at,Castaldi:2021es,Galiffi:2022tp}, and absorbers \cite{Li:2021ts}.  Recent reviews and perspectives on this rapidly advancing research area can be found in Refs. \cite{Galiffi:2022po,Pacheco:2022tv}.
While there are inherent constraints and technological challenges in the temporal modulation of constitutive parameters \cite{Hayran:2022hb}, experimental studies are continuing to progress, and recent results have demonstrated feasibility in this area \cite{Moussa:2022oo,Wang:2022mb,Liu:2022pa}.

Of special interest for this study are the recent results on {\em anisotropic} temporal metamaterials, featuring abrupt temporal switching from isotropic to anisotropic responses. These include, for instance,
temporal aiming \cite{Pacheco:2020ta} and Brewster angle \cite{Pacheco:2021te}, spatiotemporal isotropic-to-anisotropic meta-atoms \cite{Pacheco:2021si},
complete polarization conversion \cite{Xu:2021cp}, nonreciprocity and Faraday rotation \cite{Li:2022na}, and spin-temporal interactions \cite{Mostafa:2022st}.

Here, we investigate the capabilities of anisotropic temporal platforms to attain spin-controlled field manipulations. Specifically, within the recently proposed framework of {\em short-pulsed} temporal metamaterials \cite{Rizza:2022sp,Castaldi:2022ma}, we show that anisotropy can be leveraged to attain spin-dependent  analog computing on an impinging wavepacket, and to overcome some limitations inherent of isotropic scenarios. Additionally, we demonstrate that typical spin-orbit interaction effects observed in spatial, anisotropic scenarios \cite{Ciattoni:2017ev} can be translated to the temporal case, enabling efficient vortex generation.
These findings highlight the potential of temporal anisotropic metamaterials for advanced field manipulations, with diverse and wide-ranging possible applications to communications, optical processing and quantum technologies.

%%%%%%%%%%%%%%%%%%%%%%%%%%%%%%%%%%%%%%
\section{Results and discussion}
%%%%%%%%%%%%%%%%%%%%%%%%%%%%%%%%%%%%%%

%############################################################
%                Figure1
%
\begin{figure}
	\centering
	\includegraphics[width=.6\linewidth]{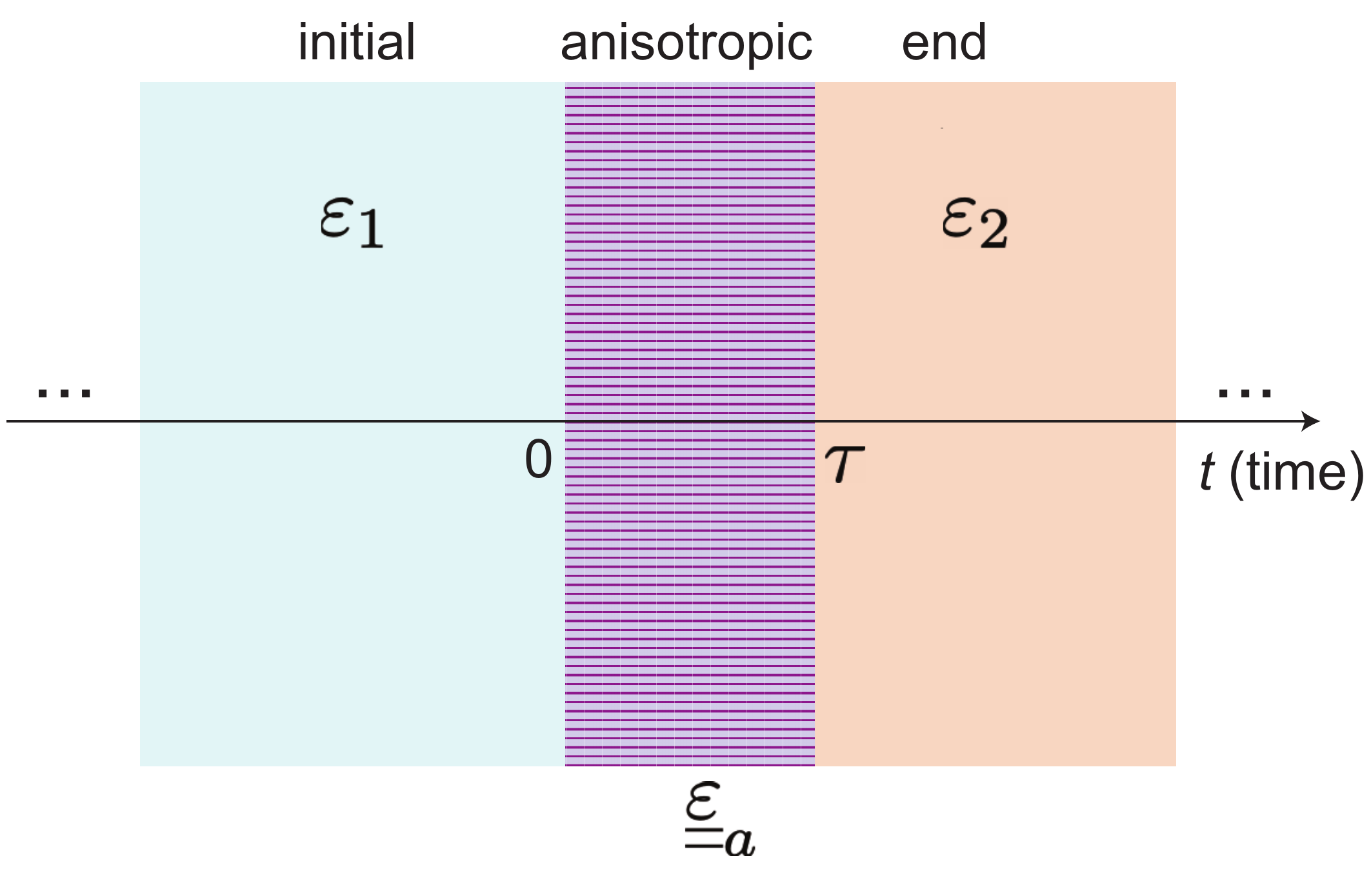}
	\caption{Schematic representation of an anisotropic temporal slab (details in the text).}
	\label{Figure1}
\end{figure}
%############################################################

%========================================================
\subsection{Problem schematic and statement}
%========================================================
As schematically illustrated in Figure \ref{Figure1}, we consider a  spatially unbounded, nonmagnetic medium described by the
the constitutive relationships 
\beq
	\label{cost_equa}
	{\bf D}=\varepsilon_0 \underline{\underline \varepsilon}(t)\cdot {\bf E}, 
	\quad {\bf B}=\mu_0  {\bf H}, 
\eeq 
which relate the electric and magnetic inductions (${\bf D}$ and ${\bf B}$, respectively) to the corresponding fields (${\bf E}$ and ${\bf H}$, respectively). In Equations (\ref{cost_equa}), $\varepsilon_0$ and $\mu_0$ are the vacuum dielectric permittivity and magnetic permeability, respectively, and 
\begin{eqnarray}
	\label{ep_equa}
	\underline{\underline \varepsilon}   (t)=
	\left \{ \begin{array}{ll}
		\varepsilon_1 \underline{\underline I}, \quad t< 0, \\
		\underline{\underline {\varepsilon}}_a=\varepsilon_{\perp} \left({\hat {\bf e}}_x\otimes{\hat {\bf e}}_x+ 
		{\hat {\bf e}}_y\otimes{\hat {\bf e}}_y\right)+\varepsilon_{\parallel} {\hat {\bf e}}_z\otimes{\hat {\bf e}}_z, \quad 0 <t <\tau, \\
		\varepsilon_2 \underline{\underline I}, \quad t > \tau,
	\end{array}	\right.
\end{eqnarray} 
is a time-varying relative-permittivity tensor, with $\varepsilon_{\nu}$ ($\nu=1,2,\perp,\parallel$) denoting real-valued constants, ${\underline {\underline I}}$ the identity tensor, and $\underline{\underline {\varepsilon}}_a$ a uniaxial tensor. Here and henceforth, ${\hat {\bf e}}_{\alpha}$ indicates an $\alpha$-directed unit vector, and $\otimes$ the dyadic product.

Equation (\ref{ep_equa}) describes a time-varying medium which, at time $t=0$, undergoes an abrupt transition from a stationary, isotropic state to a uniaxially anisotropic response; subsequently, at the time instant $t=\tau$, the response is abruptly switched back to an isotropic state, which is then maintained indefinitely. In analogy with previous studies \cite{Xu:2021cp}, we refer to the configuration above as an ``anisotropic temporal slab'.' We assume that the EM fields experience ideal temporal boundaries, i.e., discontinuous changes of the dielectric permittivity at $t=0$ and $t=\tau$. However, in our full-wave simulations (see the Methods section \ref{sec:fw}) we take into account {\em finite} rising/falling times (much shorter than the temporal slab duration $\tau$). Moreover,
as in previous studies on this subject \cite{Pacheco:2020ta,Pacheco:2021te,Pacheco:2021si,Xu:2021cp}, we assume to be far away from any material resonance, so as to neglect temporal dispersion. Considering the impact of dispersion is technically possible, but would necessitate a more sophisticated approach than the transfer-matrix method used here. Recent research has indicated that sudden variations in the plasma frequency of a Lorentzian-type dispersive medium can lead to the emergence of two shifted frequencies and the need for additional boundary conditions \cite{Solis:2021tv}. This extension will be addressed in future studies. 

From Maxwell’s equations, the EM field dynamics can be described by the vector wave equation for the electric induction, viz.,
\begin{eqnarray}
	\label{D_equa}
	\frac{\partial^2 {\bf D} }{\partial t^2}+ c^2 \nabla \times \nabla \times \left[{\underline {\underline \varepsilon}^{-1}(t)\cdot \bf D} \right]=0,
\end{eqnarray}
where $^{-1}$ denotes the inverse operator, and $c$ is the wavespeed in vacuum. In what follows, we will derive a general analytical solution, which will be subsequently particularized to two scenarios of interest.

%========================================================
\subsection{General Theory}
%========================================================
We start by considering plane-wave solutions of the type ${\bf D}({\bf r},t)=\mbox{Re} \left[{\bf d} ({\bf k},t) e^{i {\bf k} \cdot {\bf r}}\right]$, with $i$ denoting the imaginary unit, ${\bf r}=x{\hat {\bf e}}_x+y{\hat {\bf e}}_y+z{\hat {\bf e}}_z$ and ${\bf k}=k_x{\hat {\bf e}}_x+k_y{\hat {\bf e}}_y+k_z{\hat {\bf e}}_z$ the position and wave vectors, respectively, and ${\bf d}({\bf k},t)$ a time-dependent plane-wave spectrum.  From Equations (\ref{cost_equa}) and (\ref{ep_equa}), together with the Maxwell's curl equation $\nabla \times {\bf E}=-\partial_t {\bf B}$, we identify the eigenwaves of the temporal anisotropic slab, i.e., the ordinary ($s$-polarized) and extraordinary ($p$-polarized) plane waves, with eigenfrequencies
\begin{eqnarray}
	\label{om_equa}
	\omega_p=c \sqrt{\frac{k_{\perp}^2}{n_{\parallel}^2} + \frac{k_z^2}{n_{\perp}^2}} ,\quad
	\omega_s=\frac{c}{n_{\perp}} k,
\end{eqnarray}
and polarization unit vectors 
\begin{subequations}
\begin{eqnarray}
	\hat{\bf e}_p&=&\frac{k_x k_z }{k k_{\perp}} \hat{\bf e}_x +\frac{k_y k_z }{k k_{\perp}} \hat{\bf e}_y - \frac{k_{\perp} }{k} \hat{\bf e}_z,\\
	\hat{\bf e}_s&=&-\frac{k_y}{k_{\perp}}  \hat{\bf e}_x +\frac{k_x}{k_{\perp}} \hat{\bf e}_y,
\end{eqnarray}
	\label{vec_equa}
 \end{subequations}
where $n_{\perp}=\sqrt{\varepsilon_{\perp}}$, $n_{\perp}=\sqrt{\varepsilon_{\parallel}}$, $k=\left|{\bf k}\right|$, and $k_\perp=\left|{\bf k}_\perp \right|=\sqrt{k_x^2+k_y^2}$. Equations (\ref{vec_equa}) clearly hold for $k_{\perp} \neq 0$; for the case $k_{\perp} = 0$, we define instead 
\begin{eqnarray}
	\hat{\bf e}_p=\hat{\bf e}_x, \quad
	\hat{\bf e}_s=\hat{\bf e}_y.
 \label{vec_equa2}
\end{eqnarray}

Next, we investigate the scattering of a time-harmonic plane wave, which, for $t<0$, can be written as ${{\bf D}^{(i)}}({\bf r},t)=\mbox{Re} \left[{{\bf d}^{(i)}} ({\bf k}) e^{i \left({\bf k} \cdot {\bf r}- \omega_1 t\right)}\right]$.  The interaction with the anisotropic temporal slab described by Equations (\ref{ep_equa}) will generate forward (transmitted) and  backward (reflected) waves, which, for $t>\tau$, can be written as
 ${{\bf D}^{(t)}}({\bf r},t)=\mbox{Re}\left\{{{\bf d}^{(t)}} ({\bf k}) e^{i \left[{\bf k} \cdot {\bf r}-\omega_2  (t-\tau)\right]}\right\}$ and ${{\bf D}^{(r)}}({\bf r},t)=\mbox{Re}\left\{{{\bf d}^{(r)}} ({\bf k}) e^{i \left[{\bf k} \cdot {\bf r}+\omega_2  (t-\tau)\right]}\right\}$, respectively, with $\omega_{1,2}=c k/n_{1,2}$ and $n_{1,2}=\sqrt{\varepsilon_{1,2}}$ denoting the angular frequencies and refractive indices, respectively, in the initial/final medium. By enforcing the conventional temporal boundary conditions (i.e., continuity of the electric and magnetic inductions at $t=0,\tau$) \cite{Xiao:2014ra}, we can obtain the temporal transmission and reflection matrices connecting the transmitted  [${{\bf d}^{(t)}}$] and  reflected [${{\bf d}^{(r)}}$]  vector amplitudes of the electric induction in the  polarization basis of Equations (\ref{vec_equa}), viz., ${{\bf d}^{(t)}}=\underline{\underline{T}}\cdot{{\bf d}^{(i)}}$, ${{\bf d}^{(r)}}=\underline{\underline{R}}\cdot{{\bf d}^{(i)}}$, where 
\begin{subequations}
\begin{eqnarray}
	\underline{\underline{T}}({\bf k})&=&T_{pp}({\bf k}) {\hat {\bf e}}_p\otimes {\hat {\bf e}}_p+T_{ss}({\bf k}) {\hat {\bf e}}_s\otimes {\hat {\bf e}}_s, \\
	\underline{\underline{R}}({\bf k})&=&R_{pp}({\bf k}) {\hat {\bf e}}_p\otimes {\hat {\bf e}}_p+R_{ss} ({\bf k}) {\hat {\bf e}}_s\otimes {\hat {\bf e}}_s,      
\end{eqnarray}
	\label{T0_R0}
\end{subequations}
and ${\bf d}^{(j)}=d_p^{(j)} {\hat {\bf e}}_p+d_s^{(j)} {\hat {\bf e}}_s$, with $j=i,r,t$. The expressions of the scattering coefficients ($T_{pp}$,$R_{pp}$,$T_{ss}$,$R_{ss}$) are reported in the Methods section \ref{sec:am}, together with the possible generalization  to an arbitrary polarization basis.

%========================================================
\subsection{Representative examples}
%========================================================
\label{sec:RE}
%----------------------------------------------------------------------------
\subsubsection{Short-pulsed regime: Spin-dependent analog computing}
%----------------------------------------------------------------------------
\label{sec:SP}
As a leading first example, we consider the  {\em short-pulsed} regime $\tau \ll \Delta t$, with $\Delta t$ denoting a characteristic timescale of the wave dynamics. In our previous studies on {\em isotropic} configurations \cite{Rizza:2022sp}, we have shown that such regime may be interpreted as a {\em nonlocal} temporal boundary, whose response can be harnessed so as to perform elementary analog computing (e.g., derivatives) on an impinging wavepacket. Here, we explore to what extent {\em anisotropy} can be leveraged to attain {\em spin-dependent} operations.

To this aim, we assume circularly polarized plane waves propagating along the $x$-axis (i.e., ${\bf k}=k_x \hat {\bf{e}}_x$), and label with the subscripts ``+'' and ``-'' the associated spin, corresponding to the unit vectors $\hat{\bf e}_{\pm}=(-\hat{\bf e}_z\pm i \hat{\bf e}_y)/\sqrt{2}$, i.e., left- or right-handed circular (LHC or RHC) polarization, respectively, for the assumed incidence direction. 
In this case, it can be shown (see the Methods section \ref{sec:am} for details) that the relevant transmission and reflection coefficients can be approximated as 
\begin{subequations}
\begin{eqnarray}
	\label{Tpp}
	T_{++}(k_x)&=&T_{--}(k_x) \simeq \frac{1}{2} \left(1+\frac{n_2}{n_1} \right) -i \frac{\pi}{2} \left(\frac{2}{n_1}+\frac{n_2}{n_{\perp}^2}+\frac{n_2}{n_{\parallel}^2} \right)  \frac{k_x}{K}+ 
	{\cal O}\left(\frac{k_x^2}{K^2}\right),  \\
 \label{Tpm}
	T_{+-}(k_x)&=&T_{-+}(k_x) \simeq i \frac{\pi}{2} n_2 \left(\frac{1}{n_{\perp}^2}-\frac{1}{n_{\parallel}^2} \right) \frac{k_x}{K}+ {\cal O}\left(\frac{k_x^2}{K^2}\right),
\end{eqnarray}
and
\begin{eqnarray}
	\label{Rpp}
	R_{++}(k_x)&=&R_{--}(k_x) \simeq \frac{1}{2} \left(1-\frac{n_2}{n_1} \right) 
	-i \frac{\pi}{2} \left(\frac{2}{n_1}-\frac{n_2}{n_{\perp}^2}-\frac{n_2}{n_{||}^2} \right)  \frac{k_x}{K}+ 
	{\cal O}\left(\frac{k_x^2}{K^2}\right), \\
 \label{Rpm}
	R_{+-}(k_x)&=&R_{-+}(k_x) \simeq -i \frac{ \pi}{2} n_2 \left(\frac{1}{n_{\perp}^2}-\frac{1}{n_{\parallel}^2} \right) \frac{k_x}{K}+ {\cal O}\left(\frac{k_x^2}{K^2}\right),
\end{eqnarray}
\label{eq:TTRR}
\end{subequations}
where $K=2 \pi/(c \tau)$ and  ${\cal O}$ is the Landau symbol. Here and henceforth, $T_{\pm \pm}$, $R_{\pm \pm}$ ($T_{\mp \pm}$, $R_{\mp \pm}$) denote the co-polar (cross-polar) transmission and  reflection coefficients, respectively. 
%############################################################
%                Figure2
%
\begin{figure}
	\centering
	\includegraphics[width=.7\linewidth]{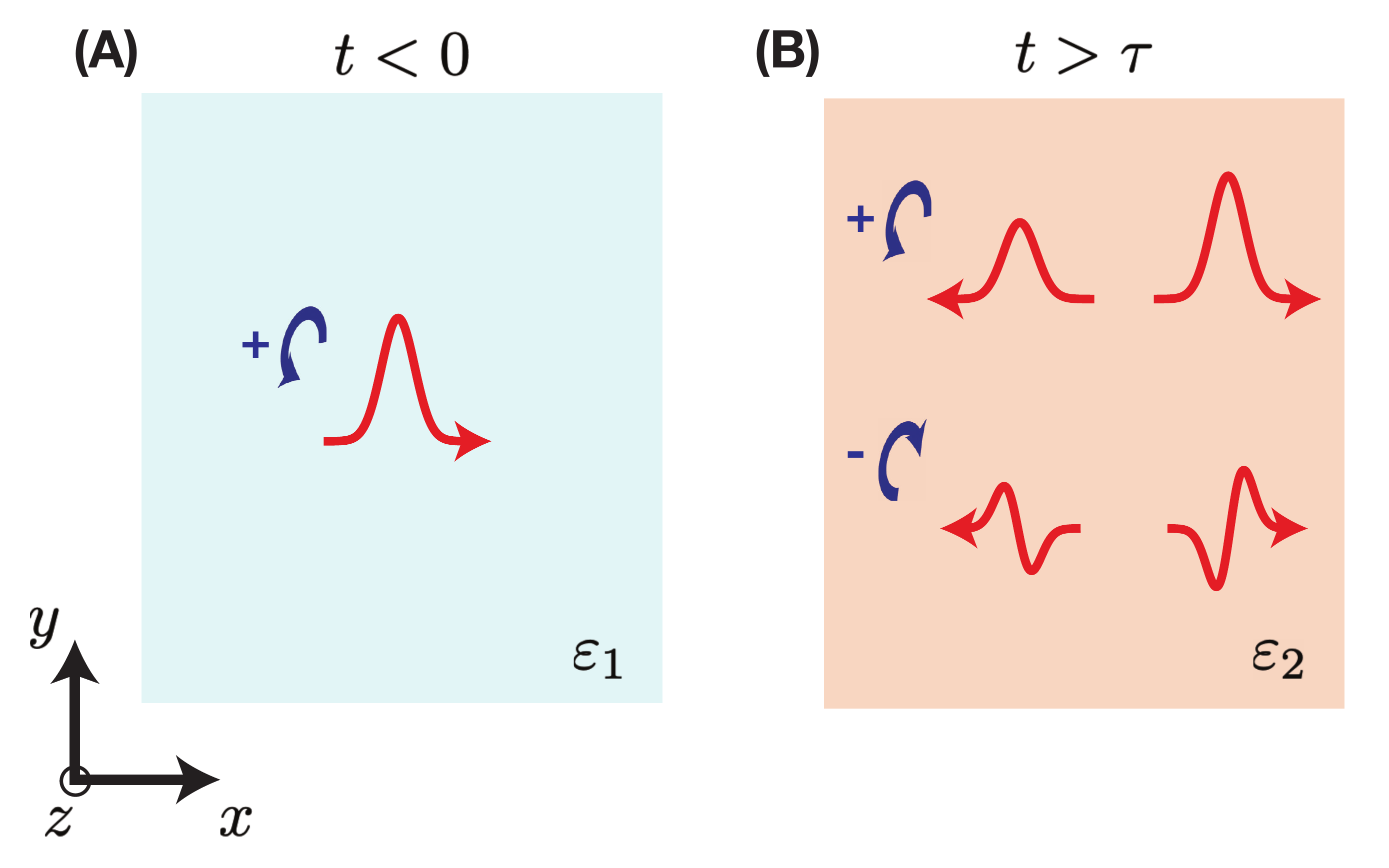}
	\caption{Schematic illustration of spin-dependent analog computing. (A) A wavepacket with a positive spin (i.e., LHC) impinges in the initial medium. (B) After the short-pulsed anisotropic temporal modulation ($t>\tau$), the co-polarized reflected (backward) and transmitted (forward) wavepacket exhibit the same profile as the impinging one, whereas the cross-polarized ones are proportional to its first derivative.}
	\label{Figure2}
\end{figure}
%############################################################

Equations (\ref{eq:TTRR}) represent the first main results of our study, since they clearly illustrate the impact of {\em local} contributions (i.e., constant terms)  and {\em nonlocal} terms (i.e., proportional to $k_x$) and their interactions with the wave spin. Remarkably, these results suggest how to exploit temporal anisotropy for designing unconventional spin-dependent analog computations. For instance, it is evident that the co-polar coefficients ($T_{\pm \pm}$, $R_{\pm \pm}$) are generally dominated by local terms, whereas the leading terms in the cross-polar ones ($T_{\mp \pm}$, $R_{\mp \pm}$) are nonlocal ($\propto ik_x$); recalling the well-known  property of the Fourier transform, these latter are amenable to first derivatives.
Therefore, as schematically illustrated in Figure \ref{Figure2}, for an impinging wavepacket with a given spin and profile [Figure \ref{Figure2}(A)],
we expect the co-polarized transmitted/reflected wavepackets to generally exhibit a similar profile, and the cross-polarized transmitted/reflected ones to be essentially proportional to its first derivative [Figure \ref{Figure2}(B)]. 

By comparison with isotropic configurations ($\varepsilon_\perp=\varepsilon_{\parallel}$) \cite{Rizza:2022sp}, where the analog-computing capabilities emerge only in the reflection (backward) response for impedance-matching conditions ($\varepsilon_1=\varepsilon_2$), we note that the anisotropy character enables (via polarization conversion) extended  operations, also in the absence of impedance matching and in the transmission (forward) response. Most important, by controlling the wave polarization, it enables elementary {\em spin-dependent} analog computing.   

For illustration and validation, we assume an incident wavepacket with a positive spin and Gaussian profile,
\beq
{\bf D}^{(i)}(x, t)=D_0 e^{-\left[\frac{x-v_1\left(t-t_s\right)}{v_1 \sigma_t}\right]^2} \hat{\bf e}_{+}, \quad t<0,
\label{eq:incD}
\eeq
with $D_0$ denoting a constant amplitude, $\sigma_t$ a characteristic timescale, $v_1=c/\sqrt{\varepsilon_1}$, and $t_s=-5 \sigma_t$.

\begin{figure}
	\centering
	\includegraphics[width=\linewidth]{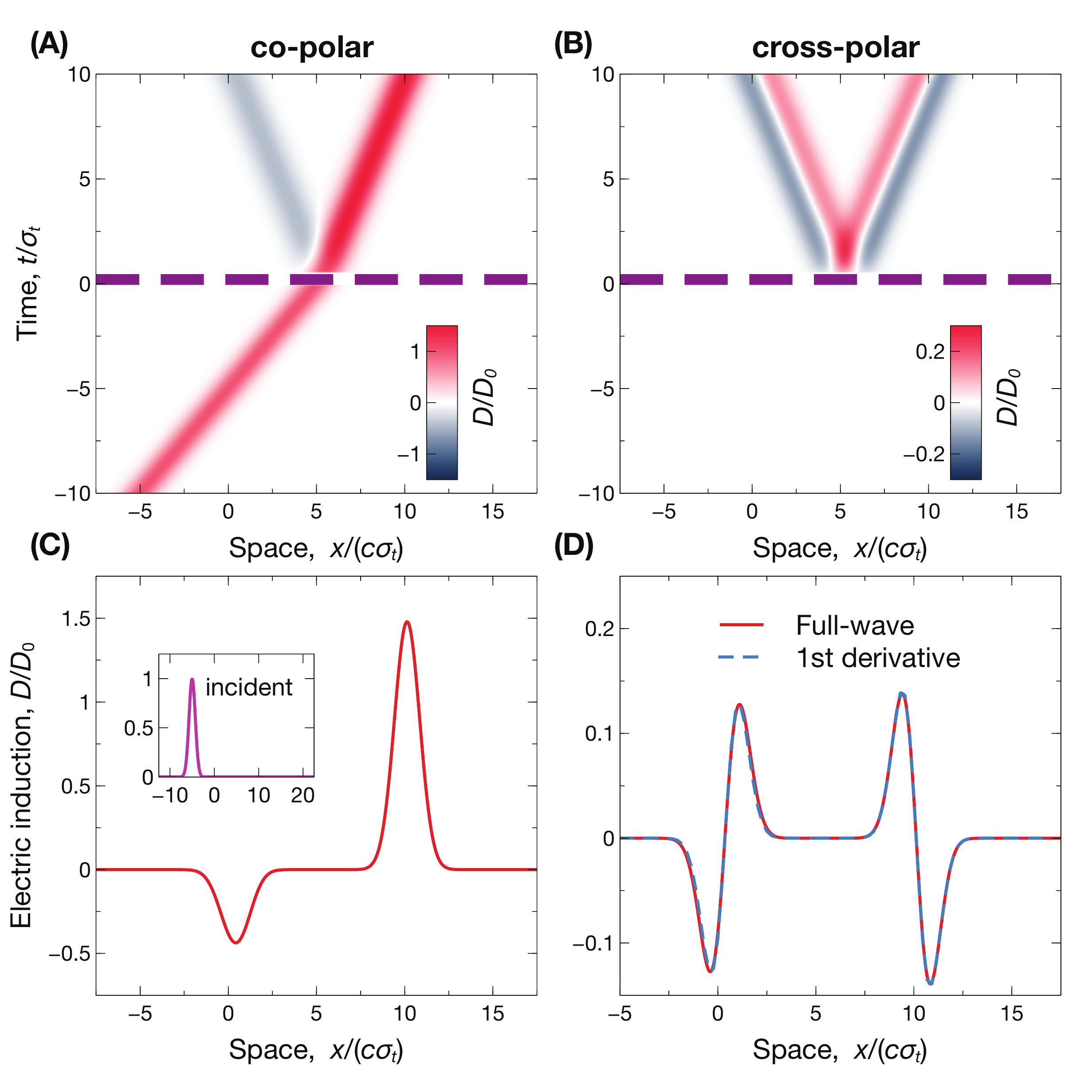}
	\caption{Example of spin-dependent analog computing. Anisotropic short-pulsed temporal slab with $\varepsilon_\perp=1$, $\varepsilon_\parallel=3$, $\varepsilon_1=1$, $\varepsilon_2=4$, and $\tau=0.5\sigma_t$, excited by the incident Gaussian wavepacket in Equation (\ref{eq:incD}) [see inset in panel (C)].   (A), (B) Space-time maps [normalized electric induction, computed from Equations (\ref{eq:TTRR})], for co-polar and cross-polar responses, respectively. The thick purple-dashed lines indicate the temporal boundaries. (C), (D) Corresponding spatial cuts at $t=10\sigma_t$, computed via full-wave simulations. The superposed blue-dashed curves indicate the expected first derivatives.}
	\label{Figure3}
\end{figure}

As a first illustrative example, we consider as initial and final relative permittivities $\varepsilon_1=1$ and $\varepsilon_2=4$, respectively, and
an anisotropic temporal slab with $\varepsilon_{\perp}=1$ and $\varepsilon_{\parallel}=4$, of duration $\tau=0.5\sigma_t$ so as to fulfill the short-pulsed assumption. Under these conditions, the co-polar transmission and reflection responses in Equations (\ref{Tpp}) and (\ref{Rpp}), respectively, are dominated by local terms, whereas the leading terms in the cross-polar responses [see Equations (\ref{Tpm}) and (\ref{Rpm})] are nonlocal ($\propto i k_x$). Figures \ref{Figure3}(A,B) show the corresponding space-time maps computed from Equations (\ref{eq:TTRR}), from which we observe the expected local reflection/transmission in the co-polar components [Figure \ref{Figure3}(A)] and the emergence (for $t>\tau$) of a cross-polar response with clearly nonlocal character [Figure \ref{Figure3}(B)]. For a more quantitative assessment, Figures \ref{Figure3}(C,D) show the corresponding spatial cuts at a fixed time instant ($t=10\sigma_t$). In the co-polar response [Figure \ref{Figure3}(C)], both the reflected (backward) and transmitted (forward) waveform are essentially scaled copies of the incident wavepacket (shown in the inset), as typically observed in conventional temporal boundaries \cite{Xiao:2014ra}. Conversely, the cross-polar responses [Figure \ref{Figure3}(D)] contain scaled copies of the first derivatives. This is the first example of an elementary analog operation that is performed only on a selected wave spin. We also notice that, by comparison with isotropic scenarios \cite{Rizza:2022sp}, now the analog-computing capabilities are enabled in transmission too, and without the need for impedance matching. This latter condition implies the possibility to perform analog computing in conjunction with frequency conversion, which is not attainable in the isotropic case \cite{Rizza:2022sp}.

\begin{figure}
	\centering
	\includegraphics[width=\linewidth]{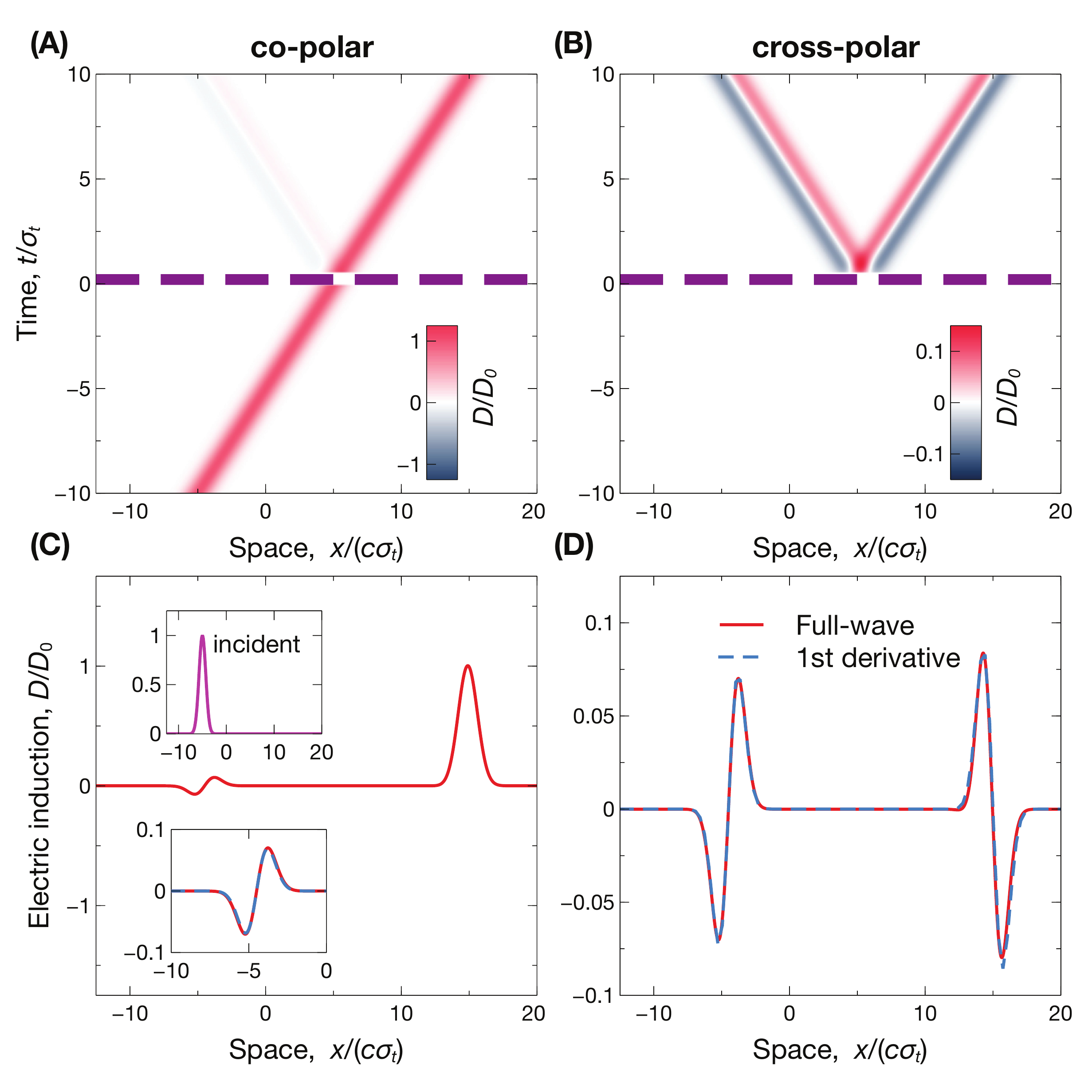}
	\caption{Example of spin-dependent analog computing. Anisotropic short-pulsed temporal slab with $\varepsilon_\perp=1$, $\varepsilon_\parallel=3$, $\varepsilon_1=\varepsilon_2=1$, and $\tau=0.5\sigma_t$, excited by the incident Gaussian wavepacket in Equation (\ref{eq:incD}) [see upper inset in panel (C)].   (A), (B) Space-time maps [normalized electric induction, computed from Equations (\ref{eq:TTRR})], for co-polar and cross-polar responses, respectively. The thick purple-dashed lines indicate the temporal boundaries. (C), (D) Corresponding spatial cuts at $t=10\sigma_t$, computed via full-wave simulations. The superposed blue-dashed curves indicate the expected first derivatives. The lower inset in panel (C) shows a magnified view of the reflection (backward) response.}
	\label{Figure4}
\end{figure}

Figure \ref{Figure4} illustrates another interesting example, where the parameters are selected so as to attain impedance matching ($\varepsilon_1=\varepsilon_2$). This implies the vanishing of the local term in the co-polar reflection response [see Equation (\ref{Rpp})], which is therefore dominated by nonlocality. As a consequence, we obtain a different combination of spin-dependent analog operations, where a first derivative is performed for both wave spins in reflection [Figures \ref{Figure4}(A,C)], and for one only in transmission [Figures \ref{Figure4}(B,D)].

\begin{figure}
	\centering
	\includegraphics[width=\linewidth]{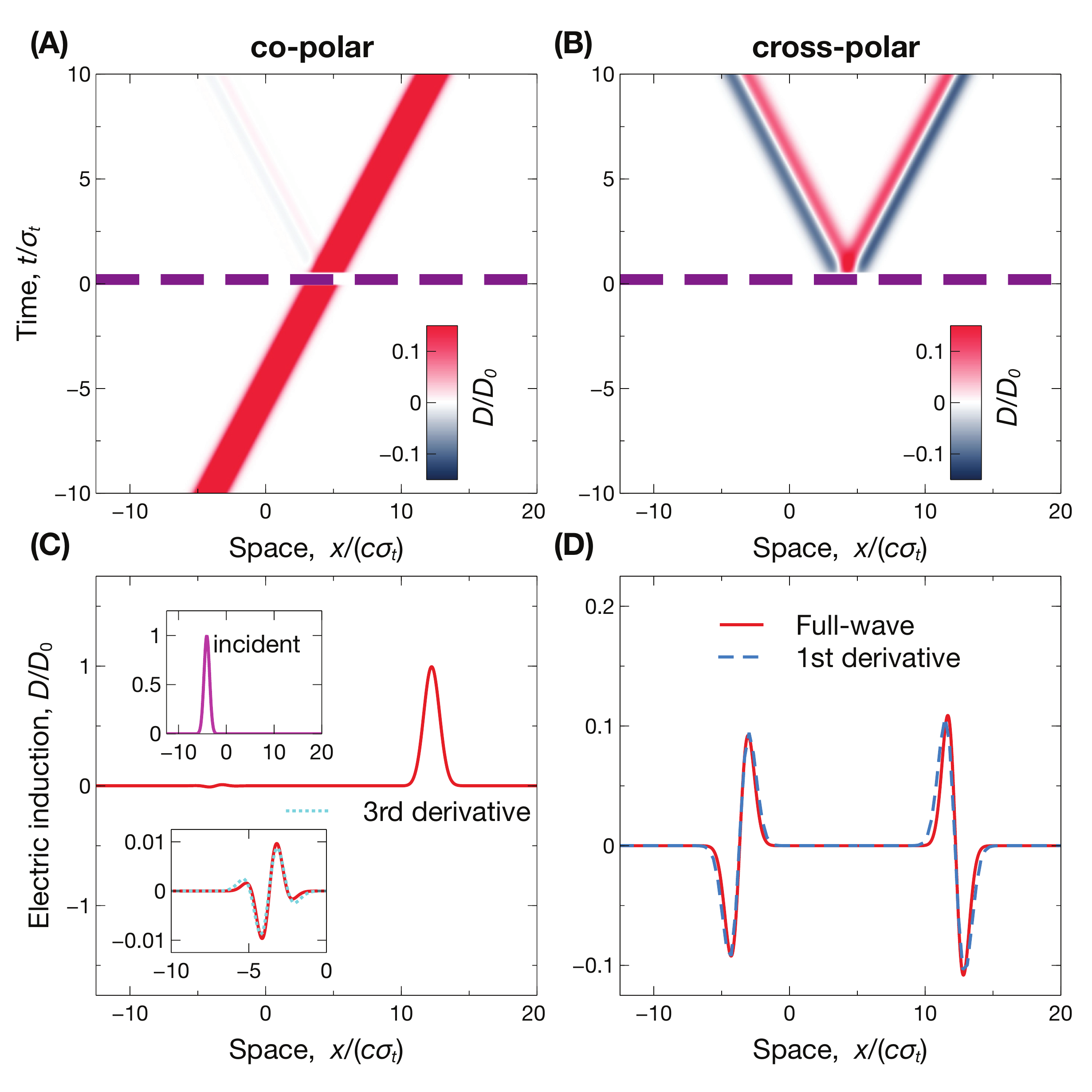}
	\caption{Example of spin-dependent analog computing. Anisotropic short-pulsed temporal slab with $\varepsilon_\perp=1$, $\varepsilon_\parallel=3$, $\varepsilon_1=\varepsilon_2=1.5$, and $\tau=0.5\sigma_t$, excited by the incident Gaussian wavepacket in Equation (\ref{eq:incD}) [see upper inset in panel (C)].   (A), (B) Space-time maps [normalized electric induction, computed from Equations (\ref{eq:TTRR})], for co-polar and cross-polar responses, respectively. The thick purple-dashed lines indicate the temporal boundaries, and the color scale in panel (A) is suitably saturated so as to show the weakest waveform. (C), (D) Corresponding spatial cuts at $t=10\sigma_t$, computed via full-wave simulations. The superposed blue-dashed and  cyan-dotted curves indicate the expected first and third derivatives, respectively. The lower inset in panel (C) shows a magnified view of the reflection (backward) response.}
	\label{Figure5}
\end{figure}

As a further variation, in Figure \ref{Figure5}, we select the parameters in such a way that both the local and first-order nonlocal terms in Equation (\ref{Rpp}) vanish. It can be shown that this condition also implies the vanishing of the second-order nonlocality, thereby leaving the third-order term ($\propto ik_x^3$) as the dominant one (see the Methods section \ref{sec:am} for details). This enables a more sophisticated response, where a first- or third-order derivative is performed in reflection, depending on the wave spin. As also observed in previous studies on short-pulsed isotropic metamaterials \cite{Rizza:2022sp,Castaldi:2022ma}, by increasing the order of the derivatives, their amplitude may decrease rapidly. However, it is worth stressing that the above parameters were merely chosen for a basic illustration of the phenomenon, and the maximization of the amplitude was not a concern; in principle, higher efficiencies may be obtained, also in view of the inherently {\em active} character of our time-varying platform. 

It is worth highlighting that more sophisticated operations can be attained by 
tailoring the short-pulsed modulation waveform  and/or
 via {\em multiple}, time-resolved short-pulsed temporal slabs, by extending to the anisotropic case of interest here the approaches developed in Refs. \cite{Rizza:2022sp,Castaldi:2022ma} for isotropic scenarios.

\begin{figure}
	\centering
	\includegraphics[width=.6\linewidth]{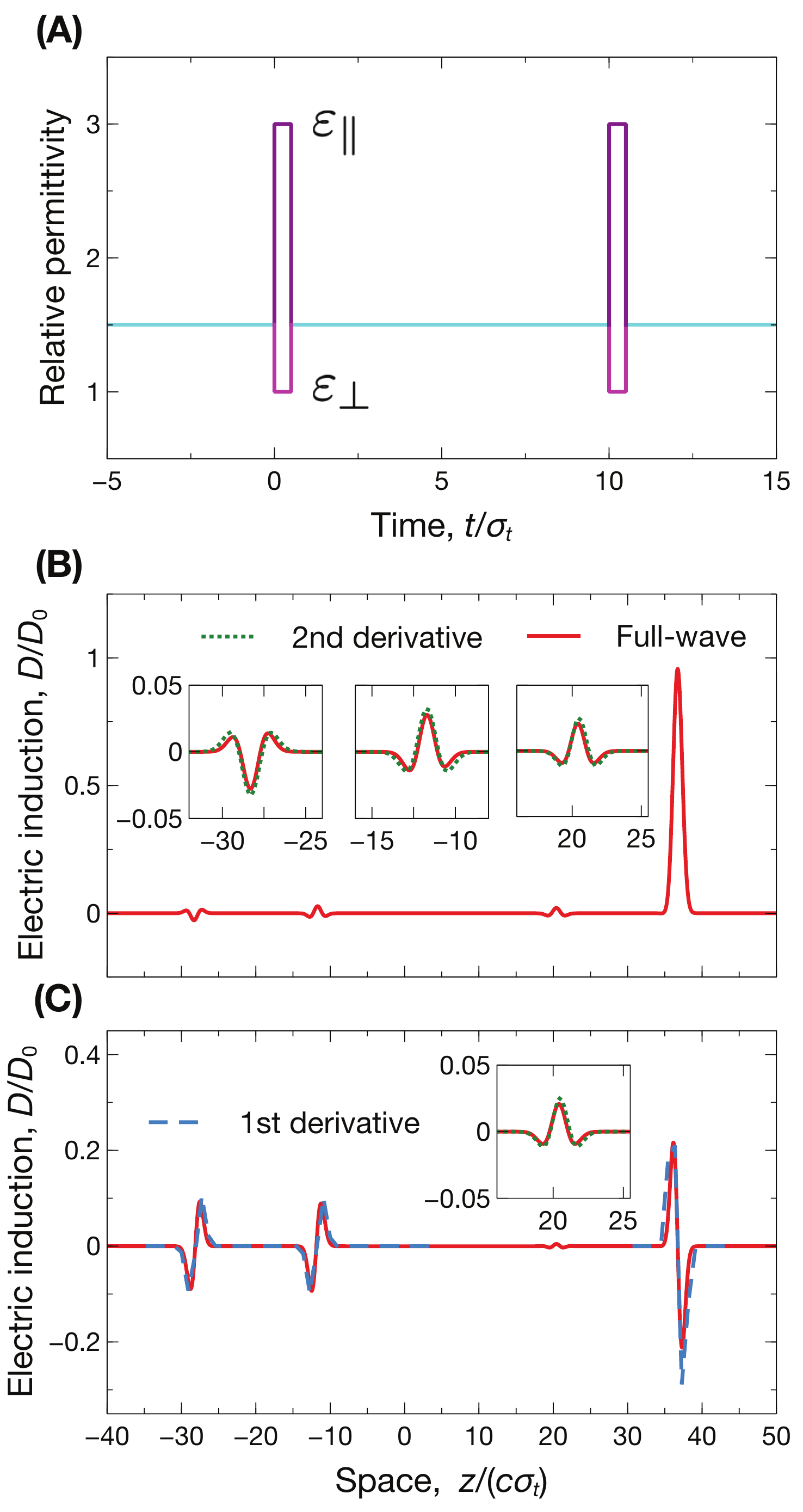}
	\caption{Example of spin-dependent analog computing. (A) Two anisotropic short-pulsed temporal slabs with $\varepsilon_\perp=1$, $\varepsilon_\parallel=3$, $\varepsilon_1=\varepsilon_2=1.5$, and $\tau=0.5\sigma_t$, starting at $t=0$ and $t=10\sigma_t$, excited by the incident Gaussian wavepacket in Equation (\ref{eq:incD}). (B), (C) Spatial cuts of normalized electric induction at $t=40\sigma_t$, computed via full-wave simulations, for co-polar and cross-polar responses, respectively. The insets show some magnified views of the responses. The superposed blue-dashed and  green-dotted curves indicate the expected first and second derivatives, respectively.}
	\label{Figure6}
\end{figure}

For a basic illustration, as shown in Figure \ref{Figure6}(A), we consider a scenario featuring two identical anisotropic short-pulsed temporal slabs with parameters as in Figure \ref{Figure5}. Similar to the isotropic case \cite{Castaldi:2022ma}, as an effect of the multiple interactions, we now observe two time-resolved waveforms in the reflection and transmission responses, with the presence of {\em composed} operations (second derivatives). However, as an effect of the anisotropy, we now obtain both the co-polar and cross-polar responses [see Figures \ref{Figure6}(B,C)]. This enables the computation of first- or second-order derivatives, depending on the wave spin.

%############################################################
%                Figure3
%
%############################################################

%############################################################
%                Figure4
%
%############################################################

%############################################################
%                Figure5
%
%############################################################

%############################################################
%                Figure6
%
%############################################################

%############################################################
%                Figure7
%
\begin{figure}
	\centering
	\includegraphics[width=.7\linewidth]{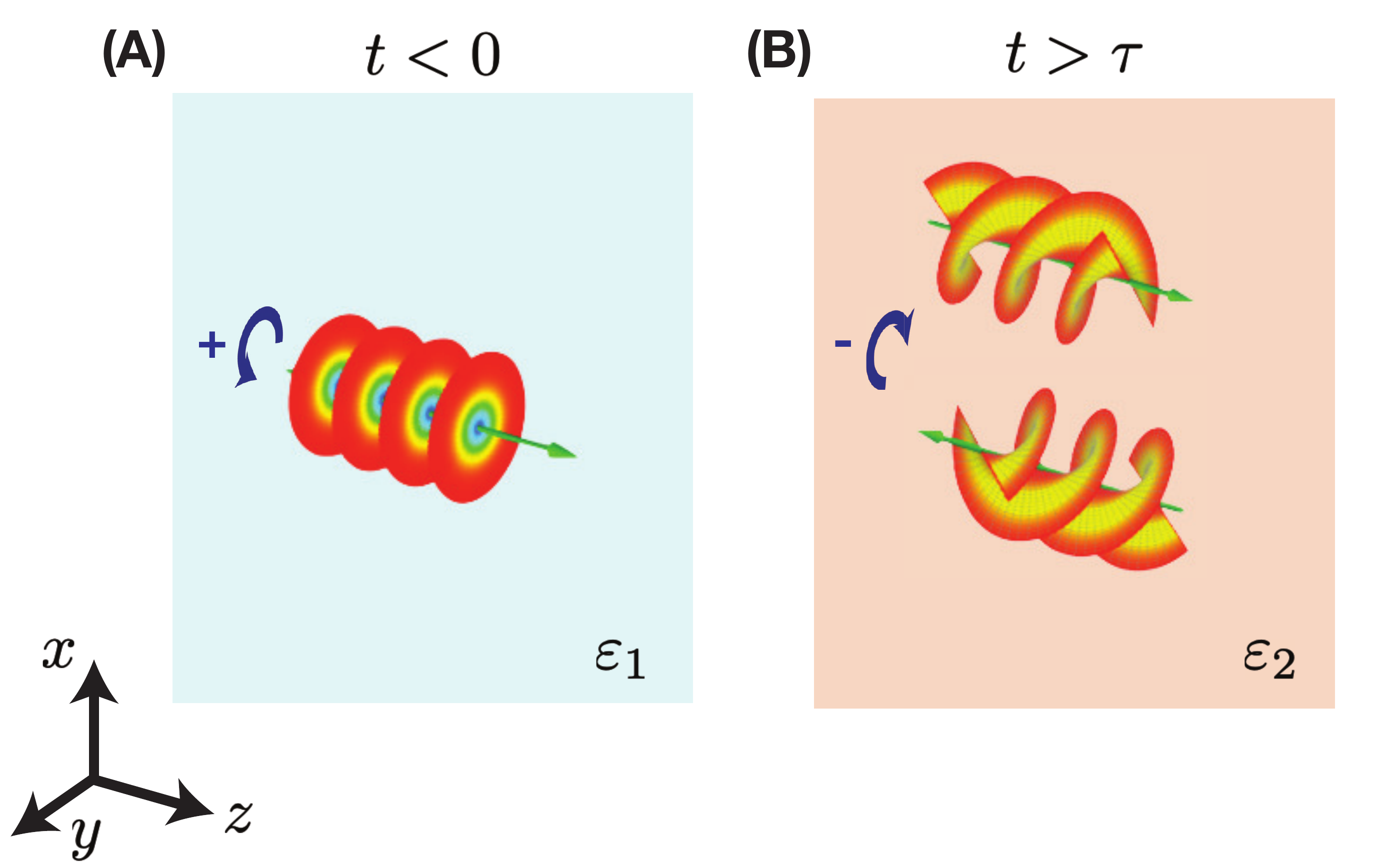}
	\caption{Schematic illustration of the spin-orbit interactions and vortex generation. (A) A circularly polarized Bessel-type beam with positive spin (i.e., LHC) and no topological charge ($\ell=0$) impinges in the initial medium. (B) After the anisotropic temporal modulation ($t>\tau$), the beam is generally converted into cross-polarized reflected (backward) and transmitted (forward) vortex beams with topological charge $\ell=2$. If the initial and final permittivities are different ($\varepsilon_1\ne \varepsilon_2$), frequency conversion (from $\omega_1$ to $\omega_2$) is attained too. Note the different incidence conditions by comparison with the scenario in Figure \ref{Figure2}.}
	\label{Figure7}
\end{figure}
%############################################################

%----------------------------------------------------------------------------
\subsubsection{Spin-orbit interactions: Vortex generation}
%----------------------------------------------------------------------------
\label{sec:SOI}
As a second representative example, we consider the generation of optical vortices. 
This phenomenon falls under the category of spin-orbit interactions of light \cite{Bliokh:2015so}, which have been extensively studied in recent years due to their potential applications in various fields. In particular, vortices have been proposed for particle trapping \cite{Gahagan:96}, optical communications \cite{Wang:11}, quantum technologies \cite{PhysRevLett.88.013601}, and microscopy imaging \cite{PhysRevE.88.033205}.

However, in temporal metamaterials, spin-orbit interactions  remain hitherto largely unexplored.
 Recently, a temporal spin-Hall effect, i.e., a spin-dependent frequency shift, has been theoretically demonstrated at a temporal boundary between bianisotropic chiral and dielectric media \cite{Mostafa:2022st}. This effect is the temporal analog of the spin-Hall effect of light, where a circularly polarized beam experiences a spin-dependent spatial shift \cite{Bliokh:2015so}. 

In conventional spatial scenarios, optical vortices can be generated through the reflection from a slab \cite{Ciattoni:2017ev} and/or the propagation along the optical axis of a homogeneous uniaxial crystal \cite{Ling:20}. In both configurations, the spin-orbit interaction effect essentially stems from the difference between the dynamics of transverse electric and magnetic fields \cite{Ciattoni:2017ev}. It appears therefore suggestive to investigate similar effects in time-varying scenarios, since temporal anisotropy likewise induces this type of asymmetry. 

To investigate optical vortex generation, we now assume a beam propagating along the optical axis (i.e., $z$) of the temporal anisotropic slab, synthesized through the superposition of plane waves with wavevector ${\bf k}_0=k_{0\perp} \left(\cos{\theta} \hat{\bf e}_x+\sin{\theta} \hat{\bf e}_y\right)+  k_{0z}\hat{\bf e}_z$, where $\theta=(0,2 \pi]$, and $k_{0\perp}$, $k_{0z}$ are pre-set wavenumbers. Accordingly, the incident beam can be written as 
\begin{subequations}
\beq
{\bf D}^{(i)}({\bf r},t)=\mbox{Re} \left[e^{i \left(k_{0z}z -\frac{c}{n_1} k_0 t\right)} {\tilde {\bf D}}^{(i)} ({\bf r}_{\perp})\right],
\eeq
where
\beq
	\label{Din}
	{\tilde {\bf D}}^{(i)} ({\bf r}_{\perp})=\int_0^{2 \pi} d \theta e^{i k_{0\perp} r_{\perp} \cos(\theta-\phi)  } \left[ d_p(\theta) \hat{\bf e}_p + d_s(\theta) \hat{\bf e}_s\right],          
\eeq
\label{eq:DDin}
\end{subequations}
$k_0=\sqrt{k_{0\perp}^2+k_{0z}^2}$, $d_{p,s}(\theta)$ is the spectral angle of the $p/s$-polarized waves, and we have assumed polar coordinates ${\bf r}_\perp = r_{\perp}(\cos{\phi} \hat{\bf e}_x+ \sin{\phi} \hat{\bf e}_y)$ in the transverse plane. Equation (\ref{Din}) is the spectral representation of a monochromatic non-diffracting wavefield, which can be expressed as a superposition of different Bessel beams \cite{PhysRevLett.58.1499}. In essence, a Bessel beam does not experience diffraction since it is the result of the superposition of plane waves exhibiting the same angular frequency and longitudinal wavenumber, and can be written as the product between a plane-wave carrier and a transverse profile.   

By exploiting the formalism in Equations (\ref{T0_R0}), we can work out analytically the expressions of the  transmitted (forward) and reflected (backward) beams (see the Methods section \ref{sec:am} for details). In particular, by
assuming the circularly polarized basis $\hat{\bf e}_{\pm}=(\hat{\bf e}_{x}\pm i \hat{\bf e}_{y})/\sqrt{2}$, the transverse components of the incident electric induction can be written as
\beq
	\label{Dt}
{\tilde {\bf D}}_{\perp}^{(i)}({\bf r}_{\perp})= \int_0^{2 \pi} d \theta e^{i k_{0\perp} r_{\perp} \cos(\theta-\phi)  } \left[ U_+^{(i)} e^{-i\theta} \hat{\bf e}_+ + U_-^{(i)} e^{i\theta} \hat{\bf e}_-        \right],   
\eeq
where we have defined $U_{\pm}^{(i)}=\left(d_p k_{0z}/k_0\mp i d_s\right)/\sqrt{2}$. Accordingly, we obtain for the
transverse components of the transmitted and reflected electric inductions
${\bf D}_{\perp}^{(t)}({\bf r},t)=\mbox{Re}\left\{e^{i \left[k_{0z}z -\frac{c}{n_2} k_0 (t-\tau)\right]} {\tilde {\bf D}}_{\perp}^{(t)} ({\bf r}_{\perp})\right\}$ and ${\bf D}_{\perp}^{(r)}({\bf r},t)=\mbox{Re}\left\{e^{i \left[k_{0z}z +\frac{c}{n_2} k_0 (t-\tau)\right]} {\tilde {\bf D}}_{\perp}^{(r)} ({\bf r}_{\perp})\right\}$, respectively, with 
\beq
{\tilde {\bf D}}_{\perp}^{(j)} ({\bf r}_{\perp})= \int_0^{2 \pi} d \theta e^{i k_{0\perp} r_{\perp} \cos(\theta-\phi)  } \left[ U_+^{(j)} e^{-i\theta} \hat{\bf e}_+ + U_-^{(j)} e^{i\theta} \hat{\bf e}_-\right],        
\eeq
and $j=r,t$. Here,
\begin{subequations}
\begin{eqnarray}
U_{\pm}^{(t)}&=&\frac{1}{\sqrt{2}}\left[T_{pp}\left(k_{0\perp},k_{0z}\right) d_p\frac{k_{0z}}{k_0}\mp i T_{ss}\left(k_{0\perp},k_{0z}\right) d_s\right],\\	
U_{\pm}^{(r)}&=&\frac{1}{\sqrt{2}}\left[R_{pp}\left(k_{0\perp},k_{0z}\right) d_p\frac{k_{0z}}{k_0}\mp i R_{ss} \left(k_{0\perp},k_{0z}\right) d_s\right],
\end{eqnarray}
\end{subequations}
with the dependence on $k_{0\perp}$ and $k_{0z}$ (rather than on ${\bf k}_0$) in the transmission/reflection coefficients being a consequence of the axial symmetry of the considered beam and anisotropic temporal slab. 

As an example, we assume that the incident beam has positive spin (LHC) and is devoid of topological charge, i.e., 
\beq
U_{+}^{(i)}=\frac{D_0}{2 \pi}  e^{i \theta},\quad U_{-}^{(i)}=0,
\label{eq:Uipm}
\eeq
with $D_0$ denoting a real-valued normalization constant; this implies that the spectral amplitudes fulfill the relationships  $d_p=k_0 D_0 e^{i \theta}/(2 \pi k_{0z} \sqrt{2} )$, $d_s=i  D_0 e^{i \theta} /(2 \pi \sqrt{2})$. 
By substituting these assumptions in Equations (\ref{Dt}), and performing the angular integration, we obtain
\begin{subequations}
\label{bessel}
\begin{eqnarray}
{\tilde {\bf D}}_{\perp}^{(i)}({\bf r}_{\perp})&=&D_0 J_0(k_{0\perp} r_{\perp}) \hat{\bf e}_+,\\
{\tilde {\bf D}}_{\perp}^{(t)}({\bf r}_{\perp})&=& D_0 \left[T_{++}\left(k_{0\perp},k_{0z}\right)J_0(k_{0\perp} r_{\perp}) \hat{\bf e}_+  - T_{-+}\left(k_{0\perp},k_{0z}\right) J_2(k_{0\perp} r_{\perp}) e^{i 2 \phi} \hat{\bf e}_-\right],\\
{\tilde {\bf D}}_{\perp}^{(r)}({\bf r}_{\perp})&=&D_0 \left[R_{++}\left(k_{0\perp},k_{0z}\right)J_0(k_{0\perp} r_{\perp}) \hat{\bf e}_+  - R_{-+}\left(k_{0\perp},k_{0z}\right)J_2(k_{0\perp} r_{\perp}) e^{i 2 \phi} \hat{\bf e}_- \right],
\end{eqnarray}
\label{Dt1}
\end{subequations}
where $J_m(\cdot)$ denote the $m$th-order Bessel functions \cite{Abramowitz:1965ho}, and $T_{\pm \pm}$, $R_{\pm \pm}$ ($T_{\mp \pm}$, $R_{\mp \pm}$) are co-polarized (cross-polarized) transmission and  reflection coefficients, respectively, for the chosen circularly polarized basis $\hat{\bf e}_{\pm}=(\hat{\bf e}_{x}\pm i \hat{\bf e}_{y})/\sqrt{2}$. 

Equations (\ref{Dt1}) clearly illustrate the spin-orbit interaction effect that can occur in an anisotropic temporal slab, with the polarization conversion giving rise to a variation of the orbital angular momentum. It appears thus possible to generate a vortex beam with topological charge $\ell=2$ in the cross-polarized reflection or transmission  response. For instance, as schematically illustrated in Figure \ref{Figure7}, it is possible to tailor the anisotropic slab parameters so that $R_{++}=T_{++}=0$ (see the Methods section \ref{sec:am} for details), i.e., so that an impinging Bessel-type beam with a positive spin and no topological charge [$\ell=0$, Figure \ref{Figure7}(A)] is converted into reflected and transmitted cross-polarized vortex beams with topological charge $\ell=2$ [Figure \ref{Figure7}(B)]. 

The above mechanism is similar to that occurring in conventional spatial scenarios involving uniaxial crystals \cite{Ling:20}, with the important difference that the temporal configuration is not necessarily bound by power conservation for EM signals, and the vortex-beam generation is also accompanied by frequency conversion if the initial and final permittivities do not coincide ($\varepsilon_1\ne \varepsilon_2$). Specifically, by calculating the time-averaged power flow of the beams before and after the temporal slab, we obtain  (see the Methods section \ref{sec:e} for further details)
\begin{equation}
\label{Pow}
\frac{P_{+}^{(t)}+P_{-}^{(t)}+P_{+}^{(r)}+P_{-}^{(r)}}{P^{(i)}_{+}}=\left(\frac{n_1}{n_2} \right)^2,
\end{equation}
where, as usual, the superscripts $i,r,t$ denote the incident, reflected and transmitted beams, respectively, whereas the subscripts $+,-$ denote positive and negative spin, respectively. Similar to the scenario of a single temporal boundary, the power conservation holds only for the impedance-matching case ($n_1=n_2$) \cite{Mai}. Furthermore, it is natural to define the vortex generation efficiency $\eta$ as the fraction of the incident power coupled with the generated vortex beam (i.e., $\eta=P_{-}^{(t)}/P^{(i)}_{+}$) \cite{Ciattoni:2017ev}. In the assumed conditions where $R_{++}=T_{++}=0$, we obtain (see the Methods section \ref{sec:e} for further details)
\begin{equation}
\label{eta}
\eta=\frac{n_1}{n_2} \left(\frac{n_1}{n_2}+|R_{-+}|^2 \right).
\end{equation} 
From Equation (\ref{eta}), it is evident that the vortex generation efficiency can become greater than one as a consequence of the lack of power conservation.   
In the impedance-matching scenario ($n_1=n_2)$, it is also possible to attain $R_{-+}=0$ (see the Methods section \ref{sec:am} for details), so that the impinging beam is perfectly converted into a cross-polarized transmitted vortex beam with unit efficiency and without frequency conversion. 

\begin{figure}
	\centering
	\includegraphics[width=.8\linewidth]{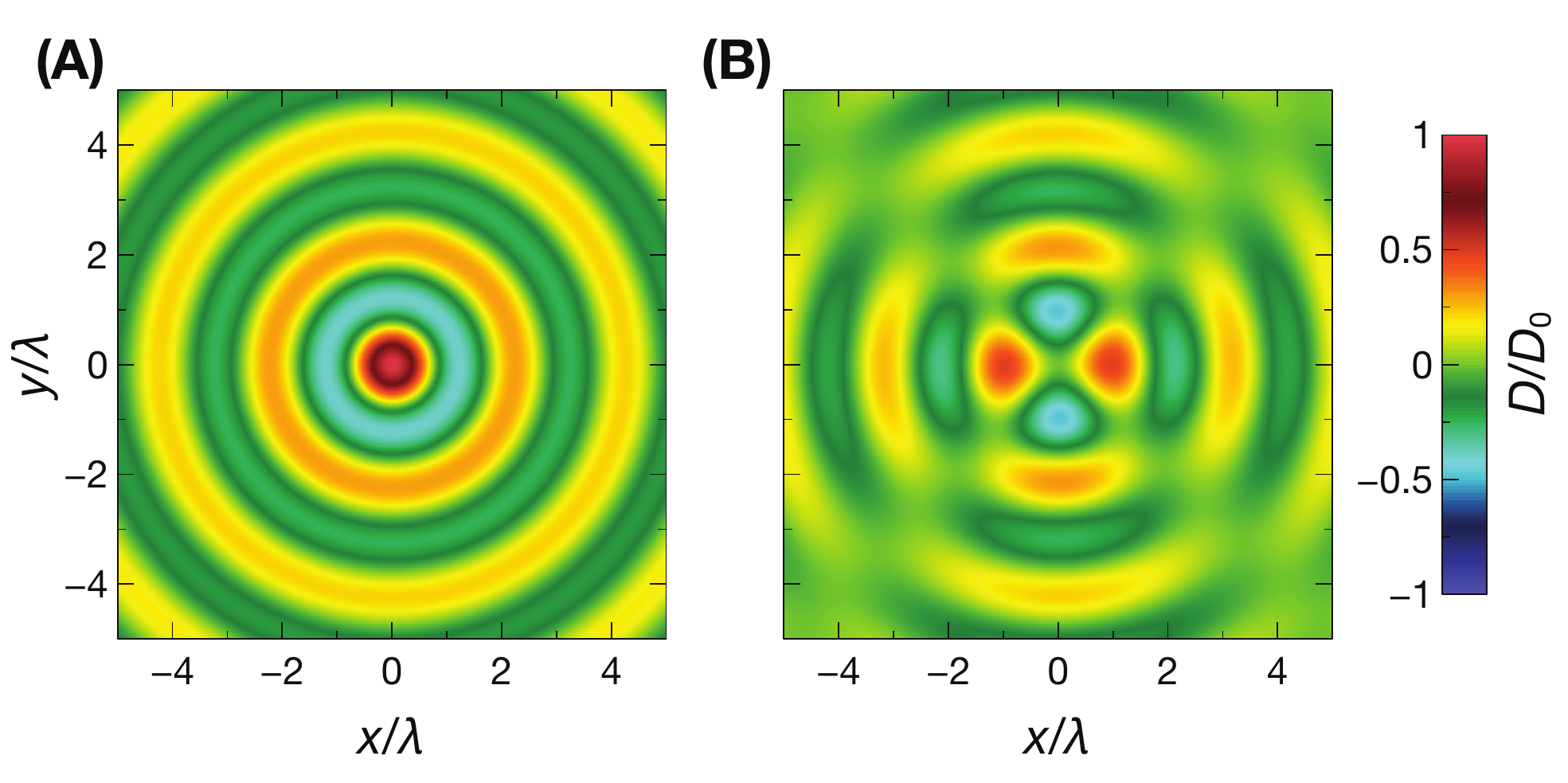}
	\caption{Example of spin-orbit interaction effects (vortex generation). Anisotropic temporal slab with $\varepsilon_{\perp}=8$, $\varepsilon_{\parallel}=1.946$, $\varepsilon_1=\varepsilon_2=1$ and $\tau=3\sqrt{2}T$, excited by a time-harmonic Bessel-type beam as in Equations (\ref{eq:DDin}), with period $T$, wavelength $\lambda=cT$, and characteristic wavenumbers $k_{0}=2\pi/\lambda$ and $k_{0\perp}=\pi/\lambda$. (A), (B) Transverse wavefronts (normalized electric induction) of impinging and transmitted (forward) cross-polarized beams, respectively,  computed via full-wave simulations at $t=-T$ and $t=19.75T$, respectively. All other (co- and cross-polarized) scattering terms are below $\sim 10^{-7}$ in the normalized scale.}
	\label{Figure8}
\end{figure}

This latter scenario is exemplified in Figure \ref{Figure8}, with parameters (given in the caption) tailored so as to perfectly convert an impinging Bessel-type beam devoid of topological charge (i.e., $\ell=0$) into a cross-polarized transmitted vortex beam with topological charge $\ell=2$.  Figures \ref{Figure8}(A,B) show the full-wave computed transverse wavefronts (at a fixed time) of the impinging and cross-polarized transmitted  beams (with all other scattering terms being below $\sim 10^{-7}$ in the normalized scale, and not shown for brevity), which confirm our theoretical predictions. 

%############################################################
%                Figure8
%
%############################################################

%%%%%%%%%%%%%%%%%%%%%%%%%%%%%%%%%%%%%%
\section{Conclusions}
%%%%%%%%%%%%%%%%%%%%%%%%%%%%%%%%%%%%%%
To sum up, we have shown that temporal anisotropy, in the form of abrupt transitions from isotropic to anisotropic dielectric permittivity (and vice versa), can be harnessed to attain spin-controlled photonic operations. These include, for instance, spin-dependent analog computing on an impinging wavepacket, and spin-orbit interaction effects for vortex generation. Overall, our outcomes indicate new pathways for advanced control of light-matter interactions, which may find potential applications in various scenarios, ranging from telecommunications to optical and quantum computing.

From the implementation viewpoint, temporal anisotropy poses technological challenges that are comparable to those encountered in isotropic temporal metamaterials (see, e.g., the discussions in Refs. \citenum{Galiffi:2022po,Pacheco:2022tv,Hayran:2022hb} and the recent experimental results in Refs. \citenum{Moussa:2022oo,Wang:2022mb,Liu:2022pa}); the additional anisotropic character could be implemented via a suitable (asymmetric) design of the reconfigurable meta-atoms, and should not pose insurmountable obstacles.

As possible extensions and follow-up studies, we are currently investigating more general short-pulsed variations (by generalizing our previous results in Ref. \citenum{Rizza:2022sp} for isotropic scenarios) as well as the combination of temporal anisotropy and conventional spatial modulations. Also of great interest are the modeling of temporal dispersion and the study of possible applications of temporally induced spin-controlled photonics to quantum technologies.

%%%%%%%%%%%%%%%%%%%%%%%%%%%%%%%%%%%%%%
\section{Methods}
%%%%%%%%%%%%%%%%%%%%%%%%%%%%%%%%%%%%%%
\label{sec:methods}

%========================================================
\subsection{Analytical modeling}
%========================================================
\label{sec:am}
For the the anisotropic temporal slab described by Equations (\ref{ep_equa}), the temporal scattering problem can be solved analytically in closed form. 
By particularizing to our scenario the approach proposed in Ref. \cite{Xu:2021cp}, the temporal reflection and transmission coefficients for the ordinary ($s$-polarized) and extraordinary ($p$-polarized) plane waves can be expressed as 
 \begin{subequations}
 \label{Tj_Rj}
 \begin{eqnarray}
 	T_{jj}({\bf k})&=&\frac{1}{2} \left[ \left(1+\frac{\omega_1}{\omega_2} \right)\cos\left(\omega_j \tau \right)-i \left(\frac{\omega_1}{\omega_j}+\frac{\omega_j}{\omega_2}\right)\sin\left(\omega_j \tau \right)         \right], \\
 	R_{jj}({\bf k})&=&\frac{1}{2} \left[ \left(1-\frac{\omega_1}{\omega_2} \right)\cos\left(\omega_j \tau \right)-i \left(\frac{\omega_1}{\omega_j}-\frac{\omega_j}{\omega_2}\right)\sin\left(\omega_j \tau \right)\right],   
  \label{eq:Rjj}
 \end{eqnarray}
\end{subequations}
 with $j=p,s$. 
 In a generic polarization basis, parameterized by the angles $\varphi$ and $\delta$,
 \begin{subequations}
 \begin{eqnarray}
 	\hat{\bf e}_1&=&\cos \varphi \hat{\bf e}_p+e^{i \delta} \sin \varphi \hat{\bf e}_s, \\ 
 	\hat{\bf e}_2&=&\sin \varphi \hat{\bf e}_p-e^{i \delta} \cos \varphi \hat{\bf e}_s,
 \end{eqnarray}
	\label{pol}
\end{subequations}
 the temporal transmission and reflection matrices can be written as
 \begin{subequations}
 \begin{eqnarray}
 	\underline{\underline{M}} =
 	\begin{pmatrix}
 		M_{11} & M_{12}  \\
 		M_{21} & M_{22}  
 	\end{pmatrix},
 \end{eqnarray}
 where
 \begin{eqnarray}
 	M_{11}&=&M_{pp}\cos^2\varphi +M_{ss} \sin^2\varphi, \\
 	M_{22}&=&M_{pp} \sin^2\varphi + M_{ss} \cos^2\varphi,\\  
 	M_{12}&=&M_{21}= \left(\frac{M_{pp}-M_{ss}}{2}\right) \sin {(2 \varphi)},
 \end{eqnarray}
\end{subequations}
 with $M=T,R$.

In Sec. \ref{sec:RE}, we have considered circular polarized waves, i.e., $\hat{\bf e}_{\pm}=(\hat{\bf e}_p \pm i \hat{\bf e}_s)/\sqrt{2}$, corresponding to $\varphi=\pi/4$ and  $\delta=\pi/2$ in Equations (\ref{pol}).  This yields,
\begin{subequations}
\label{+-}
 \begin{eqnarray}
 T_{++}({\bf k})&=&T_{--}({\bf k})=\frac{T_{pp}({\bf k})+T_{ss}({\bf k})}{2},\\
 T_{+-}({\bf k})&=&T_{-+}({\bf k})=\frac{T_{pp}({\bf k})-T_{ss}({\bf k})}{2},\\
 R_{++}({\bf k})&=&R_{--}({\bf k})=\frac{R_{pp}({\bf k})+R_{ss}({\bf k})}{2},\\
 R_{+-}({\bf k})&=&R_{-+}({\bf k})=\frac{R_{pp}({\bf k})-R_{ss}({\bf k})}{2}.
 \end{eqnarray}
 \end{subequations}
More specifically, in the example discussed in Sec. \ref{sec:SP}, we have assumed pulsed plane waves propagating along the $x$-axis, $\hat{\bf e}_{\pm}=(-\hat{\bf e}_z \pm i \hat{\bf e}_y)/\sqrt{2}$, corresponding to ${\bf k}=k_x \hat{\bf e}_x$ in  Equations (\ref{vec_equa}).    
From Equations (\ref{+-}), the expressions in Equations (\ref{eq:TTRR}) readily follow via Maclaurin series expansions up to the second order in $\tau$.

Moreover, with specific reference to the example in Figure \ref{Figure5},
the Maclaurin series expansion of the coefficient $R_{++}$ up to the third order in $\tau$ can be written as
\begin{eqnarray}
	\label{Rpp_M}
	 R_{++}(k_x) &\simeq& \frac{1}{2} \left(1-\frac{n_2}{n_1} \right) 
	-i \frac{\pi}{2} \left(\frac{2}{n_1}-\frac{n_2}{n_{\perp}^2}-\frac{n_2}{n_{||}^2} \right)  \frac{k_x}{K}
	 -\frac{\pi^2}{2}\left(1-\frac{n_2}{n_1} \right) \left(\frac{1}{n_{\perp}^2}+\frac{1}{n_{||}^2} \right) \frac{k_x^2}{K^2} \nonumber\\
& + &i \frac{\pi^3}{3} \left[ \frac{1}{n_1} \left(\frac{1}{n_{\perp}^2}+\frac{1}{n_{||}^2} \right) -n_2  \left(\frac{1}{n_{\perp}^4}+\frac{1}{n_{||}^4} \right)\right]  \frac{k_x^3}{K^3}  
 + {\cal O}\left(\frac{k_x^4}{K^4}\right).
\end{eqnarray}
From the above expression, it appears evident that the parameter choice as in Figure \ref{Figure5} implies the vanishing of the terms up to the second order, thereby leaving the third order ($\propto ik_x^3$) as the dominant one.  

In Sec. \ref{sec:SOI}, we have instead considered non-diffracting beams propagating along the $z$-axis. Accordingly, we have expressed the transverse component of the electric induction in the basis $\hat{\bf e}_{\pm}=(\hat{\bf e}_{x}\pm i \hat{\bf e}_{y})/\sqrt{2}$.  
The considered beams are a suitable superposition of plane waves with wavevector ${\bf k}_0=k_{0\perp} \left(\cos{\theta} \hat{\bf e}_x+\sin{\theta} \hat{\bf e}_y\right)+  k_{0z}\hat{\bf e}_z$, and $\theta=(0,2 \pi]$. Consequently, the temporal transmission and reflected coefficients, $T_{+\pm}\left(k_{0\perp},k_{0z}\right)$, $R_{+\pm}\left(k_{0\perp},k_{0z}\right)$ appearing in Equations (\ref{bessel}) correspond to the matrix entries reported in Equations (\ref{+-}) with ${\bf k}={\bf k}_0$; these terms do not depend on $\theta$ in view of the axial symmetry of the considered system. 

From  Equations (\ref{+-}), we observe that the vanishing of the co-polar response ($T_{++}=R_{++}=0$) can be attained by enforcing $T_{pp}=-T_{ss}$ and $R_{pp}=-R_{ss}$, which is obtained by selecting $\omega_j \tau = m_j \pi$ ($j=p,s$), with $m_p$ and $m_s$ being odd and even positive integers, respectively, or vice versa. This yields
\begin{subequations}
\begin{eqnarray}
\frac{k_{0 \perp}^2}{n_{\parallel}^2} + \frac{k_{0 z}^2}{n_{\perp}^2}&=& \left(\frac{ \pi m_p }{c \tau} \right)^2,\\
k_{0 \perp}^2+k_{0 z}^2&=&n_{\perp}^2 \left( \frac{\pi m_s }{c \tau}\right)^2,
\end{eqnarray}
\end{subequations}
which are consistent with the parameters chosen in Figure \ref{Figure8} (with $m_p=3$ and $m_s=4$).
If, as in Figure \ref{Figure8}, we also assume impedance matching, i.e., $\varepsilon_1=\varepsilon_2$, it can readily be verified from Equation (\ref{eq:Rjj}) (with $\omega_1=\omega_2$) that $R_{-+}=0$, i.e., the impinging Bessel beam is perfectly converted into a cross-polarized transmitted  vortex beam, without frequency conversion.

%========================================================
\subsection{Power flow in vortex generation}
%====================================================
\label{sec:e}
To assess the (lack of) power conservation, we evaluate the time-averaged power flowing through a disk of radius $\rho$, lying in the $z=0$
plane and centered at the origin, for the Bessel-type beams in Equations (\ref{bessel}). For the incident ($i$) and transmitted ($t$) beams, we obtain  
\begin{equation}
\label{PPP}
P_{\pm}^{(j)}=\pi \int_0^{\rho} dr_{\perp} r_{\perp} \hat{\bf e}_z \cdot \mbox{Re}\left({\bf E}_{\omega,\pm}^{(j)} \times {\bf H}^{(j)*}_{\omega,\pm} \right),
\end{equation}
where ${\bf E}_{\pm}^{(j)}=\mbox{Re}({\bf E}_{\omega,\pm}^{(j)} e^{-i \omega t})$, ${\bf H}_{\pm}^{(j)}=\mbox{Re}({\bf H}_{\omega,\pm}^{(j)} e^{-i \omega t}$) are the electric and magnetic fields with positive/negative spin, and $\omega=\omega_1,\omega_2$ for $j=i,t$, respectively. Similarly, the reflected power flow is 
\begin{equation}
\label{PPP1}
P_{\pm}^{(r)}=\pi \int_0^{\rho} dr_{\perp} r_{\perp} \hat{\bf e}_z \cdot \mbox{Re}\left({\bf E}_{\omega_2,\pm}^{(r)} \times {\bf H}^{(r)*}_{\omega_2,\pm} \right),    
\end{equation}
with ${\bf E}_{\pm}^{(r)}=\mbox{Re}({\bf E}_{\omega_2,\pm}^{(r)} e^{i \omega_2 t})$, ${\bf H}_{\pm}^{(r)}=\mbox{Re}({\bf H}_{\omega_2,\pm}^{(r)} e^{i \omega_2 t})$. By using Equations (\ref{bessel}) along with Equations (\ref{cost_equa}) and Maxwell's curl equation $\nabla \times {\bf E}=-\partial_t {\bf B}$, we obtain 
\begin{subequations}
\begin{eqnarray}
\label{PPP}
\frac{P_{\pm}^{(t)}}{P_{+}^{(i)}}&=& \frac{n_1}{n_2} \frac{\xi_2}{\xi_1} |T_{\pm +}\left(k_{0\perp},k_{0z}\right)|^2,  \\
\frac{P_{\pm}^{(r)}}{P_{+}^{(i)}}&=&  -\frac{n_1}{n_2} \frac{\xi_2}{\xi_1} |R_{\pm +}\left(k_{0\perp},k_{0z}\right)|^2, 
\end{eqnarray}
\end{subequations}
where $\xi_m=\int_0^{\rho} d r_{\perp} r_{\perp} J_{m}^2 (k_{0\perp} r_{\perp})$ with $m=0,2$. Furthermore, by exploiting Equations (\ref{Tj_Rj}) and ({\ref{+-}), it is straightforward to verify that 
\begin{equation}
\label{coeff_en}
|T_{++}|^2+|T_{-+}|^2-|R_{++}|^2-|R_{-+}|^2=\frac{n_1}{n_2}.  
\end{equation}
Recalling that  $\xi_2/\xi_1\rightarrow 1$ in the limit $\rho \rightarrow \infty$, and by combining Equations (\ref{PPP}) with Equations (\ref{coeff_en}), we obtain Equation (\ref{Pow}) reported in Section \ref{sec:SOI}. From Equation (\ref{PPP}), we obtain for the vortex efficiency 
\begin{equation}
\label{eta_S}
\eta=\lim_{\rho \to +\infty}\frac{P_{-}^{(t)}}{P_{+}^{(i)}}=\frac{n_1}{n_2}|T_{- +}\left(k_{0\perp},k_{0z}\right)|^2,
\end{equation}
which, recalling Equation (\ref{coeff_en}) with $T_{++}=R_{++}=0$, becomes Equation (\ref{eta}) reported in Section \ref{sec:SOI}.

%========================================================
\subsection{Full-wave solution}
%========================================================
\label{sec:fw}
Our full-wave solution is obtained via a rigorous numerical approach that generalizes the method introduced in Refs. \cite{Rizza:2022ne,Rizza:2022sp} for isotropic scenarios. In essence, the method operates on the spatially algebrized Maxwell's curl equations
\begin{subequations}
\begin{eqnarray}
i\frac{{\bf k}}{\varepsilon_0}\times \left[{\underline {\underline \varepsilon}}^{-1}\left(t\right)\cdot{\bf d}\left({\bf k},t\right)\right]&=&-\frac{d {\bf b}\left({\bf k},t\right)}{dt},\\
i{\bf k}\times{\bf b}\left({\bf k},t\right)&=&\mu_0\frac{d {\bf d}\left({\bf k},t\right)}{dt},
\end{eqnarray}
\label{eq:Maxt}
\end{subequations}
i.e., a system of six ordinary differential equations in time, where the wavevector ${\bf k}$ is considered as a fixed parameter. For given initial conditions and wavevector, the above system is solved numerically by means of the \texttt{NDSolve} routine available in Mathematica$^{TM}$ \cite{Mathematica}, which applies adaptively several numerical methods (e.g., Runge-Kutta, predictor-corrector, implicit backward differentiation). In our implementation, we rely on default settings and parameters. Moreover, to favor numerical convergence and to assess the effects of non-ideal temporal boundaries, we model the abrupt changes in the permittivity by means of an analytical, smooth unit-step function $U_s(t)=[\tanh(t/T_s)+1]/2$, with $T_s = 10^{-4}T$.  

For a given impinging wavefield [Equation (\ref{eq:incD}) for the example in Sec. \ref{sec:SP}, and Equations (\ref{eq:DDin}) for the example in Sec. \ref{sec:SOI}], we compute a time-dependent plane-wave spectrum, from which we derive a set of discretized wavevectors and corresponding initial conditions. For each wavevector, we then solve the system in Equations (\ref{eq:Maxt}). From these numerical solutions, we finally synthesize the physical observable of interest (electric induction) in terms of a plane-wave spectral integral, numerically implemented via fast-Fourier-transform by means of the \texttt{Fourier} routine available in Mathematica$^{TM}$ \cite{Mathematica}.

\subparagraph*{Research funding}
G. C. and V. G. acknowledge partial support from the University of Sannio via the FRA 2021 Program.

%\bibliographystyle{ieeetr}
%\bibliography{A-TMTM}

\end{document}